# A Heuristic Reputation Based System to Detect Spam activities in a Social Networking Platform, HRSSSNP

**Manoj Rameshchandra Thakur**
Computer Science Department, VJTI, Mumbai, India
manoj.thakur66@gmail.com

**Sugata Sanyal**
Advisor to the Corporate Technology Office
Tata Consultancy Services
sugata.sanyal@tcs.com

**Abstract:** The introduction of the social networking platform has drastically affected the way individuals interact. Even though most of the effects have been positive, there exist some serious threats associated with the interactions on a social networking website. A considerable proportion of the crimes that occur are initiated through a social networking platform [5]. Almost 33% of the crimes on the internet are initiated through a social networking website [5]. Moreover activities like spam messages create unnecessary traffic and might affect the user base of a social networking platform. As a result preventing interactions with malicious intent and spam activities becomes crucial. This work attempts to detect the same in a social networking platform by considering a social network as a weighted graph wherein each node, which represents an individual in the social network, stores activities of other nodes with respect to itself in an optimized format which is referred to as localized data-set. The weights associated with the edges in the graph represent the trust relationship between profiles. The weights of the edges along with the localized data-set is used to infer whether nodes in the social network are compromised and are performing spam or malicious activities.

*Keywords:* Spam; social graph; collaborative filtering; weighted graph; localized data-set, trust level .

## 1. Introduction

### 1.1 Related Work:

A considerable amount of work has been done in the area of spam detection and trust based recommendation systems for social networking platforms. A brief overview of these is as follows:

[15] suggests a dynamic personalized recommendation system, that is based on the trust between agents. It uses the concept of feedback centrality and overcomes some of the limitations of earlier recommendation systems that use other trust metrics. In the model suggested in [3] an agent tries to filter interactions based on the information that it gains from it's own social network. The model suggested in [3] identifies the impact of factors like preference heterogeneity of agents, network density among agents, and knowledge sparseness which are crucial factors for the performance of the model. The technique suggested in paper is however different from the earlier two works, in that it makes use of a weighted social graph[2] to view the relationship between profiles in a social networking platform . The technique suggested in [4] suggests a reputation based intrusion detection system to detect malicious and compromised nodes in a mobile ad-hoc network. Even though this work is not directly related to social

networking platforms, the approach suggested in [4] is relevant to the problem of filtering malicious and spam conversations among agents in a social networking platform.

Intrusion detection, which involves trying to identify malicious and compromised nodes in a given network, is similar to the process of identifying compromised agents is a given social graph representing the way in which individuals are connected. [11] discusses some of the security issues associated with distributed computing infrastructures most of which apply to a social network as well. Approaches like the ones suggested in [7][9][10][12] are instrumental in not only addressing the problem of intrusion by malicious nodes in a network but are also indirectly helpful in devising similar approaches for spam and malicious agent detection in a social network.

### 1.2 The Social Graph:

A **social graph** may be defined as a graph that represents the way individual are related to each other on the internet [16]. Even though it represents the relationship between individuals it doesn't manifest in any way the trust level among individuals. Two individuals might be related but might not have a high trust level. The ability to represent the trust level in a social graph can impart a powerful tool to detect and prevent unwanted interaction. For example, if A is not related to B wants to interact. B will try to obtain relevant information from an individual C with whom B has a high trust level and based on the inputs B will decide whether to allow A to interact or not. The suggested approach attempts to derive this trust level among profiles based on previous interactions and the relationship type and represent the trust level as weights corresponding to the edges that represent the relationship between profiles.

### 1.3 Collaborative filtering and unwanted/malicious activity detection

In a given social networking platform, the following holds, 'If Profile A is victimized by a malicious interaction by C then the chances of profile B being victimized by profile C is high'. It is this relationship that has inspired the use of collaborative filtering [1] for the suggested approach. The suggested approach adds an additional constraint that in order for profile B to detect whether profile C is trying to initiate a malicious interaction it will only try to take recommendation from profiles which it trusts ie. profiles with high trust level.

### 1.4 Weighted social graph

The suggested approach views the social networking platform as a weighted social graph [2]. Each node represents a profile (an individual), an edge represents a relationship between profiles and the weight corresponding to the edge represents the trust level among profiles. Each profile has a localized dataset associated with it that holds a table with the following format:

Profile Id (say X) → <incoming activity with X> : <outgoing activity with X>

The overall view of the weighted social graph can be depicted as shown in Figure 1.

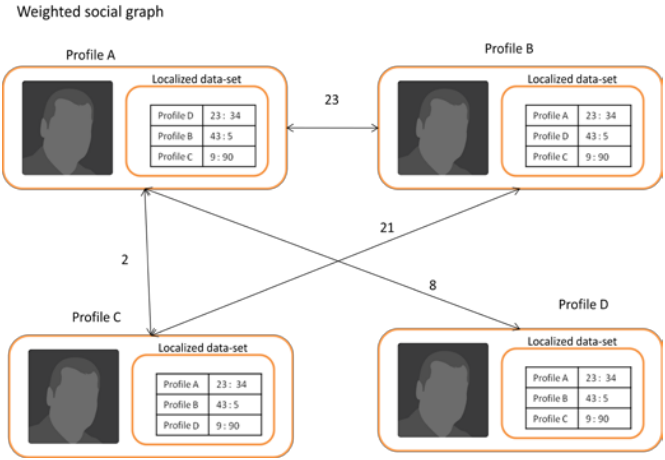

**Figure1. Weighted social graph**

## 2. Localized data-set and Edge weight calculation:

Corresponding to each profile, the localized data-set refers to a table with the following entries:

1) The first column represents profile Ids to which the given profile is connected ie. has a relationship.
2) The second column has entries in the following format:
   *Profile Id (say X)* → *<incoming activity with X> : <outgoing activity with X>*

*Incoming activity:* it represents the activities which were initiated by X in which the considered profile was the destination. These interactions include activities like message sent from X to the considered profile, friend request from X to the considered profile, comments on a photo of the considered profile etc. It must be noted that the type of interaction as mentioned earlier may vary based on the social networking website considered.

*Outgoing activity:* it represents the activities which were initiated by the considered profile wherein profile X was the destination. These interactions include activities like message sent to X, friend request sent to X, comments on a photo of profile X etc. It must be noted that the type of interaction as mentioned earlier may vary based on the social networking website considered.

In real life interactions, the trust among individuals increases over time as the interactions among the individuals increase. These interactions are however bi-directional ie. they include interactions which are initiated by both the involved individuals. The suggested approach employs this concept to calculate the trust levels among profiles. The trust level of profile X for a connected profile Y is considered is represented by *T(X, Y)* and *T(Y, X)* vice versa.

$$T(X, Y) = [O(Y)/I(Y)]$$

In order to calculate *O(Y)* and *I(Y)*, the entries in the localized data-set of X are considered corresponding to profile Y. A value for *[O(Y)/I(Y)]* that is close to 1 represents a high trust level since there. It must be noted that suggested technique will allow first few spam messages, if any, after which as the value *T(X,Y)* reduces the spam interactions will be blocked. Consider the scenario where say a fake profile A is created which sends out a friend request to a legitimate profile B. Now in such a scenario for B, $O(A)$ and $I(A)$ are both 1, since A initiated an outgoing interaction and B replied to it. However after the first few spam messages the value *[O(A) /I(A)]* will reduce thus preventing A form initiating any further spam messages.

## 2.1 Collaborative filtering of interactions

For a given profile if an incoming interaction is initiated from a profile then the profile first checks if the profile is connected. If the source profile is connected then the interaction is accepted only if the trust level between the two profiles is higher than a predefined threshold and the localized data-set of each of the profiles is updated. If however the source profile is not connected then the destination profile tries to find the trust level of the source profile with a third profile with which the destination profile has a high trust level. For example if A tries to interact with B then B will accept the interaction only if the trust level between A and B is greater than a particular threshold. However if A is not connected to B then B tries to derive or infer the trust level from a third profile C such that the trust level between B and C is higher than the threshold and C is connected to A.

## 3. Conclusion:

The suggested technique thus address the issue of malicious and spam interactions among profiles in a social networking platform in an effective way by correlating the scenario with the interactions in the society. The use of the weighted social graph imparts the suggested technique the ability to not only view and understand the way individuals are connected in a social networking platform but also reflects the trust level among individuals which helps to filter out malicious and unwanted spam interactions. It must be noted that the suggested technique will be unable to prevent spam and malicious interaction if already existing legitimate profiles with high trust level are compromised. The solution to this problem is outside the scope of this work however a potential solution to this problem is the N/R one time password system suggested in [6]. The problems of passwords of legitimate profiles being disclosed by means of attacks like password guessing attacks can be addresses by the approach suggested in [8].

## References:

[1] Recommender System [Online] http://en.wikipedia.org/wiki/Recommender_system .
[2] Weighted Graphs [Online] http://courses.cs.vt.edu/~cs3114/Fall10/Notes/T22.WeightedGraphs.pdf .
[3] Frank Edward Walter , Stefano Battiston , Frank Schweitzer, A model of a trust-based recommendation system on a social network, Autonomous Agents and Multi-Agent Systems, v.16 n.1, p.57-74, February 2008  [doi>10.1007/s10458-007-9021-x].
[4] Animesh K Trivedi, Rajan Arora, Rishi Kapoor, Sudip Sanyal and Sugata Sanyal, "A Semi-distributed Reputation-based Intrusion Detection System for Mobile Ad hoc Networks, Journal of Information Assurance and Security (JIAS), Volume 1,Issue 4, December, 2006, pp. 265-274.
[5] Social Networking Statistics [Online], http://www.internetsafety101.org/Socialnetworkingstats.htm.

[6] Vipul Goyal, Ajith Abraham, Sugata Sanyal and Sang Yong Han, The N/R One Time Password System. Information Assurance and Security Track (IAS'05), IEEE International Conference on Information Technology:Coding and Computing (ITCC'05),USA, April, 2005. Pp. 733-738, IEEE Computer Society.

[7] R. A. Vasudevan, A. Abraham, S. Sanyal and D. P. Agrawal, Jigsaw-based Secure Data Transfer over Computer Networks, IEEE International Conference on Information Technology: Coding and Computing, 2004. (ITCC '04), Proceedings of ITCC 2004,Vol. 1, pp. 2-6, April, 2004, Las Vegas, Nevada.

[8] Vipul Goyal, Virendra Kumar, Mayank Singh, Ajith Abraham and Sugata Sanyal, CompChall: Addressing Password Guessing Attacks Information Assurance and Security Track (IAS'05), IEEE International Conference on Information Technology: Coding and Computing (ITCC'05), USA. April 2005, pp 739-744, IEEE Computer Society.

[9] Ajith Abraham, Ravi Jain, Sugata Sanyal and Sang Yong Han, SCIDS: A Soft Computing Intrusion Detection System, 6th International Workshop on Distributed Computing (IWDC-2004), A. Sen et al (Eds.). Springer Verlag, Germany, Lecture Notes in Computer Science, Vol. 3326. ISBN: 3-540-24076-4, pp. 252-257, 2004.

[10] Sugata Sanyal, Dhaval Gada, Rajat Gogri, Punit Rathod, Zalak Dedhia and Nirali Mody, Security Scheme for Distributed DoS in Mobile Ad Hoc Networks Technical Report, 2004, School of Technology & Computer Science, TIFR.

[11] Rohit Bhadauria, Sugata Sanyal; "Survey on Security Issues in Cloud Computing and Associated Mitigation Techniques" International Journal of Computer Applications, Vol. 47, No. 18, June, 2012, pp. 47-66. doi> 10.5120/7292-0578.Published by Foundation of Computer Science, New York, USA.

[12] Shantanu Pal, Sunirmal Khatua, Nabendu Chaki, Sugata Sanyal; "A New Trusted and Collaborative Agent Based Approach for Ensuring Cloud Security"; in Annals of Faculty Engineering Hunedoara International Journal of Engineering; Vol. 10, Issue 1, February, 2012. pp. 71-78. ISSN: 1584-2665.

[13] Animesh Kr Trivedi, Rishi Kapoor, Rajan Arora, Sudip Sanyal and Sugata Sanyal, RISM - Reputation Based Intrusion Detection System for Mobile Ad hoc Networks, Third International Conference on Computers and Devices for Communications, CODEC-06, pp. 234-237. Institute of Radio Physics and Electronics, University of Calcutta, December 18-20, 2006, Kolkata, India.

[14] Punit Rathod,Nirali Mody,Dhaval Gada, Rajat Gogri, Zalak Dedhia, Sugata Sanyal and Ajith Abraham, Security Scheme for Malicious Node Detection in Mobile Ad Hoc Networks, 6th International Workshop on Distributed Computing (IWDC-2004), A. Sen et al (Eds.). Springer Verlag, Germany, Lecture Notes in Computer Science, Vol.3326. ISBN: 3-540-24076-4,pp 541-542, 2004.

[15] [Walter et al., 2009] Walter, F. E., Battiston, S., Schweitzer, F.: "Personalized and dynamic trust in social networks". Proc. The third ACM Conference on Recommender Systems (RecSys '09), ACM, New York, NY(2009), 197-204.

[16] Social graph [Online], http://en.wikipedia.org/wiki/Social_graph .